\documentclass[aps,prb,twocolumn,groupedaddress]{revtex4}

\usepackage[pdftex]{graphicx}
\usepackage[labelsep=period, justification=raggedright]{caption}

\begin{document}

\title{Observation of ultra-high mobility excitons in a strain field by space- and time-resolved spectroscopy at sub-Kelvin temperatures}

\author{Yusuke Morita,$^1$ Hirosuke Suzuki,$^1$ Kosuke Yoshioka,$^2$ and Makoto Kuwata-Gonokami$^1$}

\affiliation{$^1$ Department of Physics, Graduate School of Science, The University of Tokyo, 7-3-1 Hongo, Bunkyo-ku, Tokyo 113-0033, Japan\\ $^2$  Photon Science Center, Graduate School of Engineering, The University of Tokyo, 7-3-1 Hongo, Bunkyo-ku, Tokyo 113-8656, Japan}

\date{\today}

\begin{abstract}
 We measured basic parameters such as the lifetime, mobility, and diffusion constant of trapped paraexcitons in Cu$_2$O at very low temperatures (below 1 K) using a dilution refrigerator. To obtain these parameters, we observed the space- and time-resolved luminescence spectrum of paraexcitons in strain-induced trap potential. 
We extracted the lifetime of 410 ns from the measurements of the decay of the luminescence intensity. By comparing the experimental results and numerical calculations, we found that the mobility and the diffusion constant increase as the temperature of the paraexcitons decreases below 1 K. In particular, we obtained a mobility of 5.1 $\times$ 10$^7$ cm$^2$/V$\cdot$s at the corresponding temperature of 280 mK. To the best of our knowledge, this value is the highest exciton mobility that has been measured. These results show that the mean free path of the paraexcitons reaches a size ($\sim 300\ \mathrm{\mu}$m) comparable to that of the cloud of trapped paraexcitons ($\sim 100\ \mathrm{\mu}$m).
From our analyses, we found that the spatial distribution of the paraexcitons can reach a distribution that is defined by the statistical distribution function and the shape of the three-dimensional trap potential at ultra-low temperatures (well below 1 K). Our survey shows that the ultra-low temperature ensures that the Bose--Einstein condensation transition in a trap potential can be investigated by examining the spatial distribution of the density of 1s paraexcitons.
\end{abstract}

\maketitle

\section{Introduction}
An exciton is a bound state of one electron and one hole in a solid. Electrons and holes are classed as fermions and, naively, excitons are considered bosons. Bose--Einstein condensation (BEC) is expected to occur at high densities and low temperatures.\cite{blatt1962bose,hulin1980evidence} Consideration of the BEC of photo-excited quasi particles is an illuminating problem from a modern perspective, as these particles essentially constitute a many-body system following non-equilibrium quantum statistics. 

The exciton BEC has been studied extensively since the 1960s, both experimentally and theoretically.\cite{blatt1962bose,keldysh1968collective,hulin1980evidence,casella1963possibility,snoke1987quantum,hanamura1977condensation} In particular, 1s paraexcitons in cuprous oxide (Cu$_2$O) are regarded as one of the most promising candidates for realization of the exciton BEC in bulk semiconductors.\cite{mysyrowicz1979long, hulin1980evidence} As we explain in the following section, 1s paraexcitons have an extremely long lifetime \cite{mysyrowicz1979long,yoshioka2010quantum} because they are very weakly coupled to the radiation field. Experimentally, the BEC of 1s paraexcitons has been studied intensively since the 1990s.\cite{snoke1990evidence_PRB,snoke1990evidence,lin1993bose,yoshioka2011transition,schwartz2012dynamics} However, as discussed below, the exciton BEC has not been directly observed to date, and attempts are ongoing. 

Certain temperature and density regions were targeted in earlier studies of the 1s paraexciton BEC. Initially, searches for the exciton BEC were performed in Cu$_2$O by cooling these semiconductor samples to superfluid temperatures ($\sim$2 K.) \cite{snoke1987quantum, snoke1990evidence, lin1993bose} Typically, nanosecond pulsed lasers were used to achieve a critical density of approximately $10^{17}$ cm$^{-3}$ at 2 K. However, no decisive evidence of the BEC of paraexcitons was obtained despite considerable effort, including careful line-shape analyses in time-resolved experiments and absolute luminescence intensity measurements.\cite{snoke1987quantum, snoke1990evidence, lin1993bose} The physical reason why the system did not cross the BEC phase boundary remained unknown for 20 years. In the meantime, an experimental method was developed in which the absolute exciton density could be measured using the hydrogen-like internal transitions. \cite{jorger2003infrared, kuwata2004time, jorger2005midinfrared} Application of this method revealed that the density-dependent decay of 1s paraexcitons through two-body inelastic collisions between them is an important obstacle to achieving the critical density. \cite{yoshioka2010quantum} Specifically, the effective lifetime becomes approximately 100 ps at the BEC transition density at 2 K, which is too short for 1s paraexcitons to reach a thermal equilibrium state. \cite{o1999auger, o2000relaxation} Therefore, to achieve the BEC of 1s paraexcitons, reduction of the critical density is required to increase the effective lifetime. Since the 2010s, experiments at temperatures below 1 K have begun, aiming to achieve the BEC transition density of 10$^{16}$ cm$^{-3}$ or below.\cite{yoshioka2011transition,stolz2012condensation,yoshioka2013generation}

Two unique problems affect paraexcitons in Cu$_2$O at sub-Kelvin temperatures. First, the 1s paraexciton diffusion enters a ballistic regime \cite{trauernicht1984highly,trauernicht1986drift} that hinders efficient accumulation of the 1s paraexciton density. The other problem is that the refrigerator cooling power decreases as the temperature decreases. A solution to both problems is the creation of 1s paraexcitons in a trap potential with moderate excitation power. The trap potential confines the diffusive paraexcitons to a limited spatial region, yielding a gas of dense paraexcitons with moderate excitation power. The trap potential is created by applying an inhomogeneous stress on a crystal,\cite{trauernicht1986thermodynamics,naka2004two} and the potential is nearly harmonic in three dimensions; thus, the system becomes sufficiently simple to be handled theoretically. In Ref. 12, a $^3$He-refrigerator was used to cool 1s paraexcitons to 800 mK and a phenomenon suggesting BEC transition at 10$^{16}$ cm$^{-3}$ in a trap potential was reported, although the condensate was unstable; this is called the “relaxation explosion.” To obtain direct evidence of the BEC transition and to increase the condensate fraction, dilution refrigerators are now being used, and the targets for the BEC excitons are being set to increasingly low densities and temperatures (10$^{15}$ cm$^{-3}$ at 100 mK). Note that the loss rate induced by two-body inelastic collisions is decreased at reduced BEC transition density. Recently, the energy distribution of 1s paraexcitons was confirmed to reach a classical thermal distribution at 97 mK\cite{yoshioka2013generation} and 450 mK \cite{stolz2012condensation} (those researchers claimed that the paraexciton temperature at the bottom of the trap potential is 230 mK.\cite{sobkowiak2015hydrodynamic}) Theoretical investigations for paraexciton thermalization are also reported with the assumption of certain values of the exciton-phonon scattering rate and the two-body inelastic collision rate  at ultra-low temperatures. \cite{sobkowiak2015hydrodynamic,semkat2017multicomponent}

If 1s paraexcitons can be treated as ideal Bose particles in the trap potential, the paraexciton mass, temperature, and trap potential shape are the only parameters determining the steady-state spatial distribution of the trapped exciton cloud. Moreover, condensation of distributions in both momentum space and real space is expected at the BEC transition, as for atomic BEC.\cite{davis1995bose, bradley1995evidence, anderson1995observation} However, it is not self-evident whether the spatial distribution of 1s paraexcitons reaches the spatially global thermal distribution of ideal Bose particles defined by the statistical distribution function and the three-dimensional trap potential shape. Indeed, this question has already been raised in earlier studies.\cite{trauernicht1986thermodynamics,sobkowiak2015hydrodynamic} The steady state of the spatial distribution is determined by the paraexciton transport within the paraexciton lifetime. In general, exciton transport is governed by phonon scattering at high temperatures \cite{bardeen1950deformation} and by impurity scattering and local potential fluctuations at low temperatures, for which the phonon population is very low. These scatterings prevent the exciton spatial distribution from relaxation to the spatially global thermal distribution within the limited exciton lifetime. Therefore, understanding of the transport properties of 1s paraexcitons at ultra-low temperatures is very important for experimental realization and theoretical investigation of the exciton BEC.

 To study the transport properties of paraexcitons in Cu$_2$O, space- and time-resolved measurements are required. Paraexciton transport at temperatures above 1.2 K in Cu$_2$O has been studied previously.\cite{trauernicht1984highly, trauernicht1986drift} It was confirmed both experimentally and theoretically that exciton transport is determined by the interactions between the excitons and phonons. Moreover, studies \cite{trauernicht1986drift} on 1s paraexciton transport in Cu$_2$O in a trap potential have revealed that the stress on the Cu$_2$O changes the interactions between the 1s paraexcitons and phonons, with the transport results depending on the degree of stress. However, transport of 1s paraexcitons at dilution-refrigerator temperatures has not been experimentally studied. We have no knowledge of the change in the interaction between 1s paraexcitons and phonons at sub-1-K temperatures. Furthermore, it is unclear whether the impurity scattering and local potential fluctuations become potentially effective at ultra-low temperatures.

	Therefore, this study investigates the unknown transport properties of 1s paraexcitons and the cooling dynamics from initial excitation in a strain-induced trap potential at sub-1-K temperatures. A dilution refrigerator is used to cool the lattice to 40 mK. An intensified charge-coupled detector (ICCD) camera is used to detect the space- and time-resolved luminescence from the 1s paraexcitons. The time evolutions of the transport properties and local temperature of the 1s paraexcitons are successfully tracked from the initial generation. We extract the diffusion constant and mobility at sub-1-K temperatures for the first time by comparing the results of the experiment and numerical calculations. Hence, we investigate the time required for 1s paraexcitons to reach a spatially steady state and compare the spatial distribution with the distribution expected when assuming thermal equilibrium in the trap potential.  

	Our study clarifies whether the system of 1s paraexcitons realizes the spatially global thermal distribution of ideal Bose particles defined by 100-mK temperature, the chemical potential (the number of 1s paraexcitons), and the trap potential, and whether we can treat 1s paraexcitons as a system independent of phonon scattering and impurity scattering, which are present in most cases. This work also contributes to theoretical investigation of the spatial distribution of 1s paraexcitons with the BEC transition, which may be affected by the interactions between paraexcitons and phonons, impurity scattering, or random potential.

	The remainder of this paper is organized as follows. We describe the basic properties of excitons in Cu$_2$O and our experiment setup in sections 2 and 3, respectively. We report lifetime measurement at a sub-1-K temperature in section 4, and spatio-temporal dynamics measurements of the spatially resolved luminescence spectra of 1s paraexcitons in a trap in section 5. In section 6, we present numerical calculations that reproduce the experimental results, and compare these results for the spatio-temporal dynamics in section 7, extracting the diffusion constant and mobility. Finally, we comment on the future research outlook and present our conclusions in section 8.

\section{Basic properties of excitons in cuprous oxide}
\subsection{Excitons in Cu$_2$O}
Cu$_2$O is a well-known host material for typical Wannier--Mott excitons,\cite{yu2010fundamentals} because of the large binding energy ($\sim$150 meV) of the lowest exciton state, i.e., the so-called ``yellow series exciton state.'' This is a bound state of a spin 1/2 electron in the lowest conduction band ($\Gamma_6^+$) and a hole in the highest valence band ($\Gamma_7^+$), and exhibits a hydrogen-like Rydberg energy series. The lowest exciton level (principal quantum number $n$ = 1) is split into two states: a triply degenerate orthoexciton state ($\Gamma_5^+$) and a nondegenerate paraexciton state ($\Gamma_2^+$). The paraexciton energy level is 12 meV lower than that of the orthoexcitons, and this 1s paraexciton state is the ground state of a yellow series exciton. Direct recombination of paraexcitons ($\Gamma_2^+$) is optically forbidden for all orders of transitions by the parity and spin selection rules although paraexcitons are very weakly allowed through phonon-assisted recombination processes. Further, when stress is applied to the crystal, direct recombination of paraexcitons becomes very weakly allowed because of the coupling with an exciton state from the green series with the finite strain field. The weakness of the coupling between the paraexcitons and radiation field is the source of the exceptionally long paraexciton lifetime (typically more than hundreds of nanoseconds). 
\subsection{Trap potential and transport properties of paraexcitons in Cu$_2$O}
The stress dependence of the energy levels of the 1s excitons in Cu$_2$O was studied in detail.\cite{waters1980effects,mysyrowicz1983stress} The energy of 1s paraexcitons depends on the spatially dependent strain tensor induced by the applied inhomogeneous stress. Therefore, the inhomogeneous strain creates the spatial distribution of the energy shift of the 1s paraexciton level. In particular, a specific stress application called ``Hertzian contact'' \cite{markiewicz1977strain, gourley1978spatial, trauernicht1986thermodynamics} generates a specific strain profile that acts as a three-dimensional harmonic trap potential, the shape of which can be predicted from the material properties of the round plunger, applied force, and crystal orientation.

In general, the transport properties of excitons in solids are dominated by their interactions with phonons. Therefore, the paraexciton mobility is related to the scattering rate with phonons and increases as the temperature decreases because of reduction in the phonon density. Furthermore, the energy-momentum conservation law prohibits exciton--acoustic-phonon scattering processes when the exciton momentum is below a specific value, which is determined by the dispersion relations of the excitons and phonons (so-called ``freezing-out''). 
In addition, the scattering rate between paraexcitons and phonons is strain-dependent. \cite{trauernicht1986drift} Paraexciton transport in a strain-induced harmonic trap potential at temperatures of 1.2--10 K has been studied in detail.\cite{trauernicht1986drift} In those experiments, the temperature dependence of the mobility was observed at different stress regions (ranging from 1 to 3.2 kbar). Note that paraexcitons drift to the bottom of the trap potential when they are generated at some distance from the bottom.\cite{trauernicht1986drift} The paraexciton mobility can be extracted from time-resolved measurements of this drift. 
The corresponding experiment results show that, in the lowest-stress region (1 kbar), the mobility dependence on the temperature ($T$) is proportional to $T^{-5/2}$, but proportional to $T^{-3/2}$ in the highest-stress region (3.2 kbar). These results imply that (1) paraexciton transport is determined by the interactions between the paraexcitons and longitudinal acoustic (LA) phonons at low applied stress (including zero stress) and (2) interactions between paraexcitons and transverse acoustic (TA) phonons occur when the applied stress is relatively high. 
Interaction between paraexcitons and TA phonons is prohibited on account of the symmetry requests in Cu$_2$O. However, when stress is applied to the crystal, the crystal symmetry is lowered and interaction between the paraexcitons and TA phonons is allowed. The sound velocity of TA phonons (1.3 km/s) is lower than that of LA phonons (4.5 km/s). Therefore, slow paraexcitons, which are prohibited from emitting LA phonons, can still emit TA phonons. Interactions between paraexcitons and TA phonons change the $T^{-5/2}$ dependence of the exciton mobility to $T^{-3/2}$.
In recent experiments,  TA phonons under a finite strain field have contributed critically to reduction of the exciton temperature well below 1 K.\cite{yoshioka2013generation} However, the transport properties of paraexcitons in a strain-induced trap have not been studied at such low temperatures to date. Experimental verification is required as to whether the transport properties can be predicted through extension of the known mechanism acting at temperatures exceeding 1.2 K, or whether they are influenced by the impurity scattering and local potential fluctuations. 
This work addresses this issue experimentally.

\section{Experiment setup}
\subsection{Dilution refrigerator}

The experiment setup for the lifetime and transport properties of the paraexcitons is shown in Fig. 1(b). We measured the space- and time-resolved luminescence from the paraexcitons in a trap potential at a base temperature of 40 mK. As described above, paraexcitons in Cu$_2$O are very weakly coupled to the radiation field. Therefore, to obtain sufficient luminescence signal intensity, we constructed an experiment setup that allows for efficient paraexciton generation, high detection efficiency, and long exposure time, which requires sufficient stability. Below, we describe our dilution refrigerator, sample setting, excitation configuration, and designs for paraexciton luminescence detection.

Our apparatus was based on a cryogen-free dilution refrigerator (Oxford Instruments, DR-400, employed in Ref. 20) having the following five unique features. \\
(1)	Windows were attached in four directions, providing optical access to the sample stage. We carefully chose the appropriate window material to avoid input of excess thermal radiation to the sample stage.\\
(2)	The refrigerator had a high cooling power of 400 $\mathrm{\mu}$W at the base temperature of 100 mK. 
Even with careful design for optical access, a finite amount of thermal radiation is inevitable. The high cooling power compensated for this incoming heat and facilitated realization of the lower temperature of 40 mK.\\
(3)	The cryo-free system could run continuously for a long period of time, which was an important prerequisite for the experiment. We could routinely set the exposure time for several hours depending on the signal strength.\\
(4)	We set a collective lens (focal length: 15 mm, effective aperture: 3 mm) inside the refrigerator to observe the luminescence from the trapped 1s paraexcitons. A large solid angle for emitted light collection was achieved. However, it was necessary to finely adjust the lens position inside the refrigerator to realize imaging with a high spatial resolution. We controlled the lens position using a piezo-electric motor that was also attached to the refrigerator cold sample stage.\\
(5)	The refrigerator was specially designed for reduced sample stage vibration, which was experimentally confirmed to be sufficiently small (much less than our spatial resolution of $\sim$10 $\mu$m) to facilitate high-resolution imaging. 

\subsection{Sample}
We used a Cu$_2$O crystal identical to that in Ref. 20. The sample was a naturally grown pure single crystal. 

The sample surface consisted of the [100], [011], and [01$\bar{1}$] crystal planes. The sample dimensions were 5.3 $\times$ 5.3 $\times$ 8.0 mm. The excitation beam propagated along the sample [011] axis. The collective lens was set along the sample [01$\bar{1}$] axis. Sufficient thermal contact was achieved between the mixing chamber and [100] crystal surface to maintain a very low lattice temperature under optical excitation.
An inhomogeneous stress was created along the [100] axis of the sample through Hertzian contact to generate the spatial trap potential. The Hertzian contact was established by contact between the flat surface of the sample and a lens with a spherical surface (BK7, radius of curvature = 15.57 mm), which was set inside the sample stage. The strain distribution created by the Hertzian contact directed the maximum shear strain away from the surface inside the sample, and the inhomogeneous strain field acted as the trap potential. A piezo-electric motor pushed the lens against the sample and controlled the amount of applied stress. We roughly checked the magnitude of the applied strain by observing the birefringence pattern in the crystal using a charge-coupled device (CCD) camera.

\begin{figure}
    \begin{tabular}{l}

      % 1
      \begin{minipage}{1\hsize}
       
        \includegraphics[clip, width=8.6cm ]{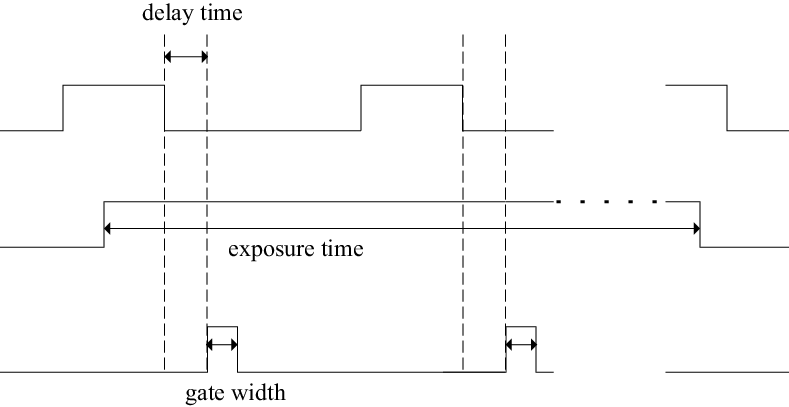}
          \hspace{1.6cm} (a)
      
      \end{minipage}\\ \\

      % 2
      \begin{minipage}{1\hsize}
        
        \includegraphics[clip, width=8.6cm ]{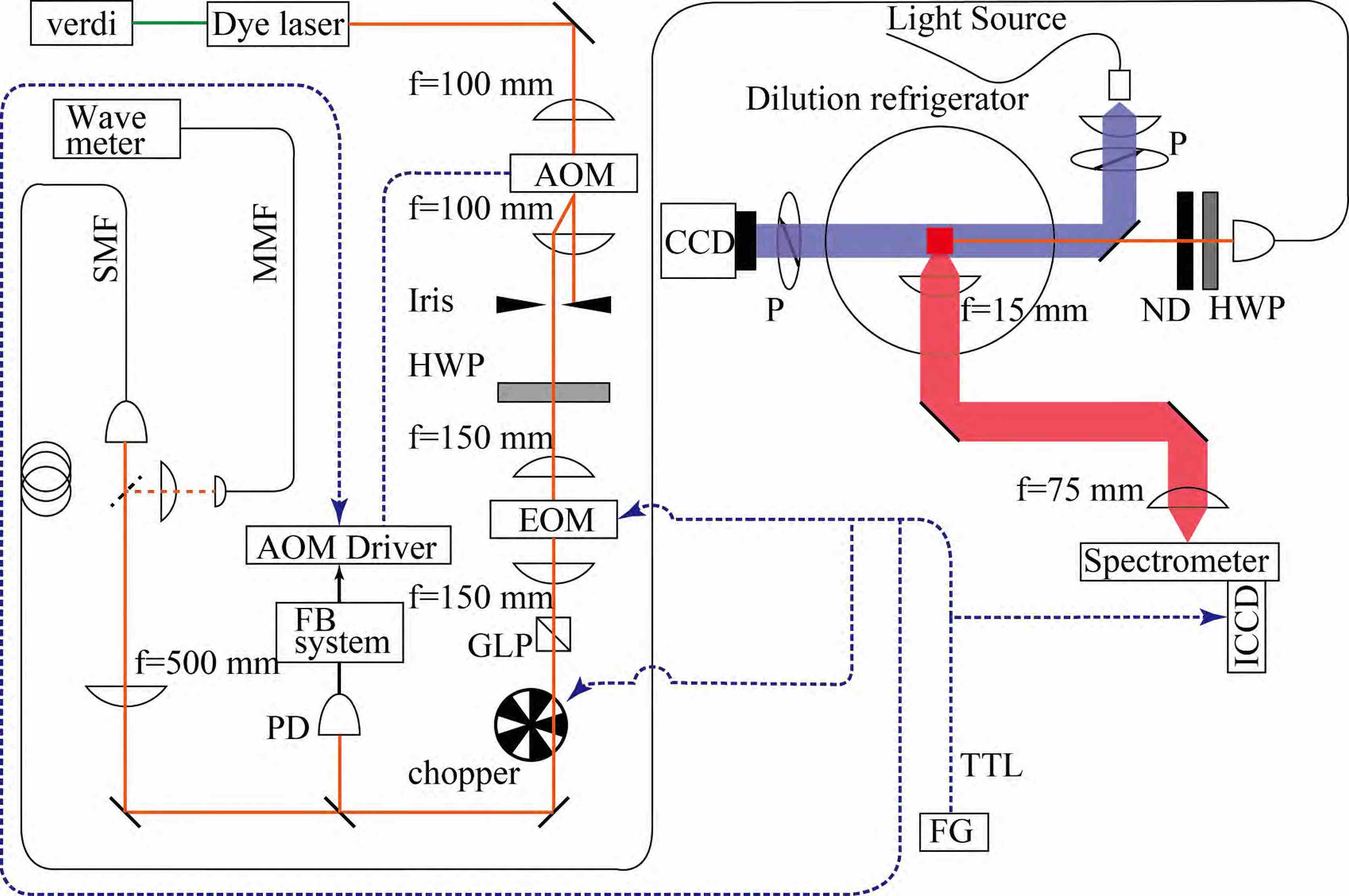}
          \hspace{1.6cm} (b)
        
      \end{minipage}

    \end{tabular}
    \caption{ (a) Timing charts. (Upper row) ON/OFF cycle of AOM driver, (middle row) ICCD exposure duration, (lower row) ICCD gate. (b) Experiment setup, with pump-laser optical path (orange solid line) and electrical paths of controlled devices (blue dotted lines).
The red and blue arrows show the optical path of the exciton luminescence and that for observation of the birefringence pattern in the crystal. SMF, MMF, FB system, PD, HWP, GLP, CCD, ND and FG: single-mode fiber, multi-mode fiber, feedback system, photodetector, half-wave plate, Glan laser polarizer, CCD camera, neutral density filter, and function generator, respectively. 
}
    \label{fig1}
  
\end{figure}

\subsection{Excitation light}
We observed the paraexciton transport in our experiment. Thus, a large number of paraexcitons were required to obtain sufficient luminescence intensity. However, as noted above, the paraexciton--radiation field coupling is very weak and, thus, direct creation of paraexcitons is inefficient. Therefore, we used the process for converting orthoexcitons to paraexcitons as an efficient means for creating paraexcitons. We generated orthoexcitons via a longitudinal-optical (LO)-phonon-assisted absorption process with a pump laser (Coherent Inc., 899-21 Ring Dye laser, Rhodamine 6G). The dye laser was pumped by a frequency-doubled Nd:YVO4 laser (Coherent Inc., Verdi-V10). We could select the dye-laser oscillation wavelength within the range of 560--620 nm. The resonant wavelength for the LO-phonon-assisted absorption varied with the position in the trap potential, because the orthoexciton energy level varied with the strain distribution. We chose a 606.7-nm wavelength to selectively generate orthoexcitons around the bottom of the trap potential. The dye-laser wavelength was measured by a wavelength meter (Bristol Instruments, 771 series laser spectrum analyzer) through a multi-mode fiber connection. The excitation beam was introduced to the sample along the [011] axis and was focused on the trap bottom through a focusing lens set outside the refrigerator.
	To perform the time-resolved experiments, it was necessary to prepare pulsed excitation light emitted from the continuous-wave (cw) laser. To simplify the numerical analysis involving step-like changes in the intensity, a buildup and intensity drop duration of the chopped excitation laser shorter than the timescale of the trap frequency and the paraexciton decay was preferable. We used the first-order diffraction of an acousto-optic modulator (AOM) to modulate the cw beam intensity. The chopping speed was determined by the AOM radio frequency and the beam size; thus, an AOM with high resonance frequency ($\sim$200 MHz, IntraAction Corp., ATM-200C1) was selected. 
In addition, the following two procedures were essential for reliable experiments. \\
(1)	We carefully constructed an AOM driving circuit to realize a high extinction ratio for the chopped excitation laser. If the extinction ratio is low, the residual light causes continuous paraexciton feeding, yielding a limited dynamic range of paraexciton density undesirable for precise lifetime measurements. To further improve the extinction ratio, we controlled the excitation laser intensity through polarization rotation using an electro-optic modulator (EOM), a half wave plate, and a Glan laser polarizer. Hence, the extinction ratio became $\sim$10$^6$.\\
(2)	We also constructed a feedback system for the AOM driving power to stabilize the excitation light intensity. The AOM driving power determined its diffraction efficiency and the excitation light intensity. Time-resolved measurements require long data acquisition time of typically 1 h. Therefore, stabilization of the excitation light intensity is important; this was monitored by a photodetector, the signal from which was received in the feedback system.

\subsection{Luminescence detection}
We detected the space- and time-resolved luminescence spectra via direct recombination of paraexcitons. We used a 50-cm imaging spectrometer (Spectra Pro-500i, Acton Research Corporation) to resolve the luminescence energy. The maximum energy resolution of the spectrometer was 50 $\mathrm{\mu}$eV. We used an intensified CCD (ICCD, iStar DH334T, Andor Technology) camera. The luminescence signal was exposed within the gate applied to the ICCD camera, allowing detection of the time- and space-resolved luminescence. The gate pulse on the ICCD camera was generated by a digital delay/pulse generator (DG535, Stanford Research Systems). The timing chart is depicted in Fig. 1(a). The timing for chopping the excitation light with the AOM was also controlled by another electrical pulse from the DG535. The repetition frequencies of these two pulses were identical. We determined the time evolution of the spatially resolved luminescence spectrum of the paraexcitons by varying the interval between these two electrical pulses. Three types of luminescence appeared in our spectral region of interest: 
(1)	Direct luminescence from orthoexcitons; 
(2)	Phonon-assisted luminescence from orthoexcitons;
(3)	Direct luminescence from paraexcitons. 
We focused on the third type.

\section{Measurement of paraexciton lifetime at sub-Kelvin temperatures}
We performed two types of experiment: the lifetime measurement and measurement of the spatio-temporal dynamics to estimate the mobility and diffusion constant of paraexcitons. As discussed in section 2, these are unknown parameters at 100 mK. We changed the excitation configurations (the excitation intensity and positions) according to the experiment type.
For the lifetime measurement, we observed the paraexciton luminescence decay after deactivating the excitation light. The following three points were important with regard to the excitation configurations.\\
(A-1)  The excitation laser intensity could not be excessively high. A strong excitation laser generates high-density paraexcitons. High paraexciton density (typically, $>\sim$10$^{14}$ cm$^{-3}$) may yield density-dependent processes such as two-body inelastic collisions, which complicate the decay processes and lifetime analyses.\\
(A-2) The excitation power should not be too weak. At low paraexciton density, the processes of trapping paraexcitons to the impurity centers may complicate the decay processes.\\
Therefore, we set the peak intensity of the quasi-cw excitation light to 300 nW for the lifetime measurements. The AOM repetition frequency was 500 Hz.\\
(B) We set the excitation position to the bottom of the trap potential. Thus, the paraexciton drift process could be ignored during the lifetime extraction.\\
The detection setup featuring the spectrometer and ICCD camera is explained here. The time-resolved intensity of the paraexciton luminescence is the only parameter required for the lifetime estimation, and was obtained by spectrally resolving the three luminescence signals listed above, as shown in Fig. 2. However, a high energy resolution was not required. Opening of the spectrometer slit decreases the energy resolution but increases the collection efficiency. Therefore, we opened the spectrometer entrance slit to 1 mm. The energy resolution was 400 $\mathrm{\mu}$eV. The energy of the paraexciton luminescence peak was approximately 2018.3 meV. We integrated the signal spectrally over 1 meV to measure the paraexciton luminescence intensity. 

\begin{figure}
 \includegraphics[width=8cm]{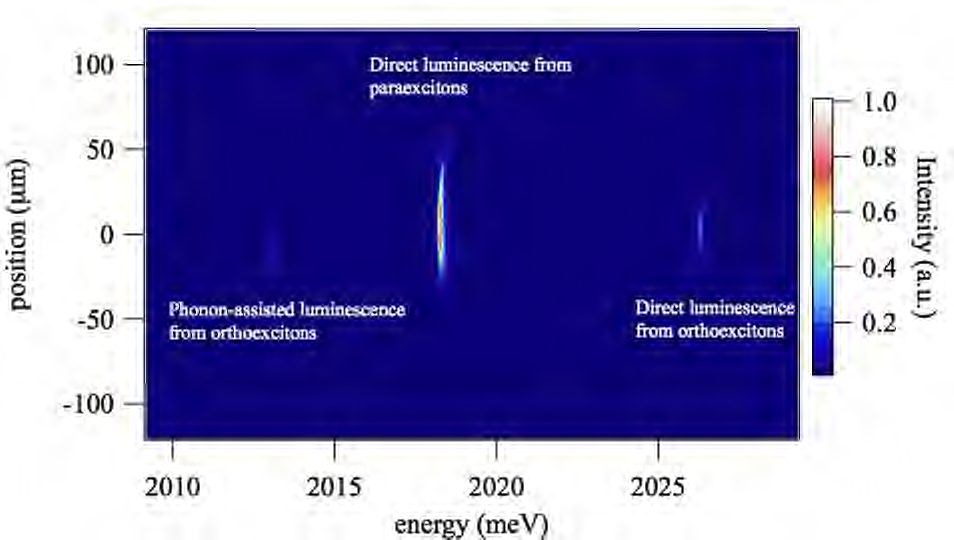}
 \caption{\label{fig2a} Typical space- and energy-resolved luminescence image at 75-mK lattice temperature. The excitation beam penetrated near the bottom of the strain-induced trap potential. Excitation power: 300 $\mathrm{\mu}$W; gate width: 100 ns; exposure time: 100 s; accumulation number: 5; and spectrometer slit width: 30 $\mu$m, yielding a higher energy resolution (50 $\mu$eV) than in the present experiment for measuring lifetime (400 $\mu$eV). The luminescence (energy $\sim$2013.0 meV) was the $\Gamma_3$-phonon-assisted orthoexciton emission. The signals at approximately 2018.3 and 2026.2 meV were the paraexciton and orthoexciton direct emissions, respectively.}
 \includegraphics[width=7.0cm]{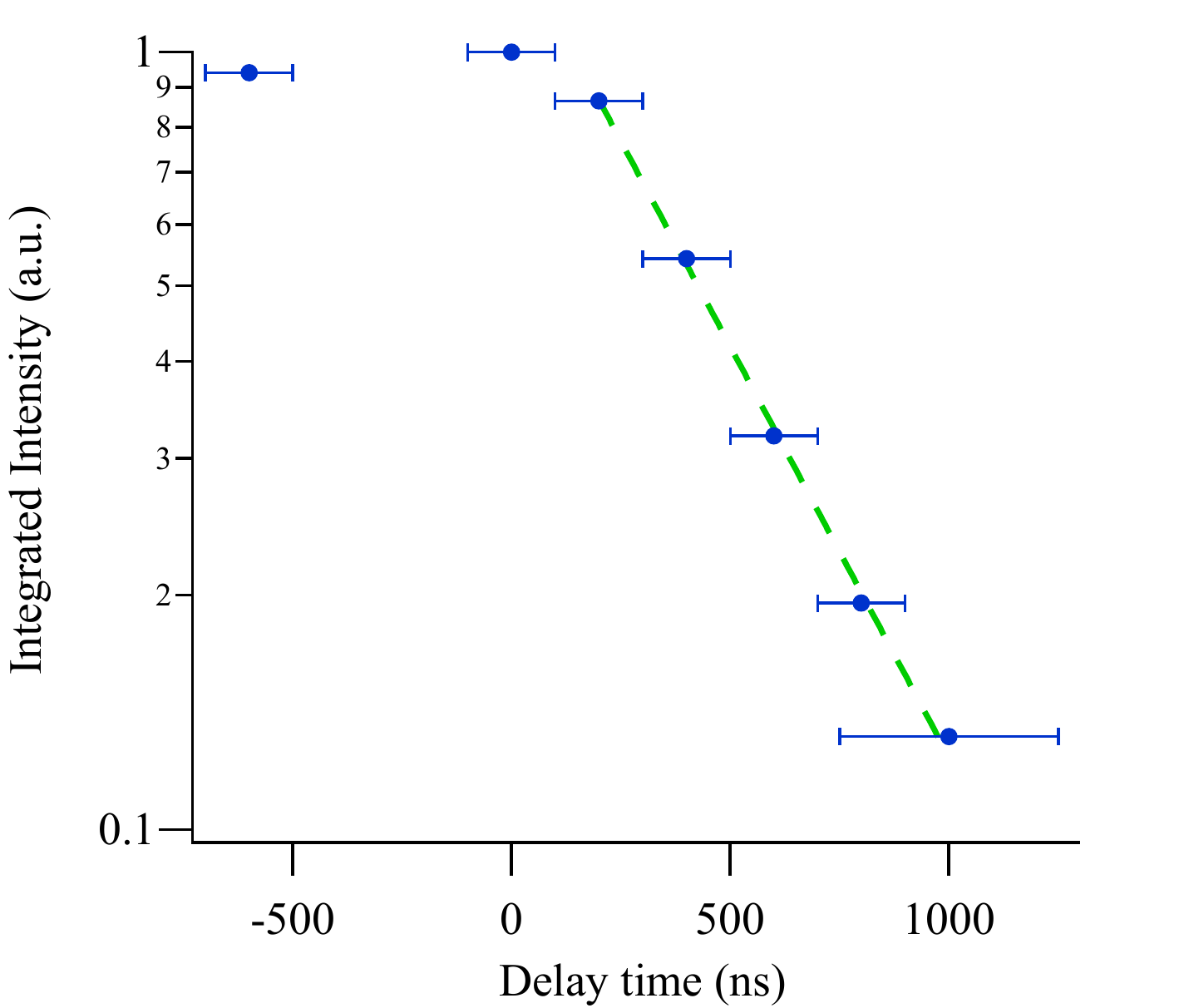}
 \caption{\label{fig2b} Time-resolved measurement results of luminescence intensity for paraexciton direct emission at 2018.3 meV (lattice temperature: 42 mK). The experiment setup and time origin measurement are explained in the text. The blue circles, horizontal bar, and green dashed line show the experiment data at each gate, the gate width, and the fitting curve, respectively. }
\end{figure}

A typical lifetime measurement result is shown in Fig. 3. The ICCD gate width was set to 200 or 500 ns. The exposure time was typically set to 1 h. The time origin was set to the point where the luminescence intensity began to decrease. We did not use the result for the gate containing the time origin for fitting, because the seemingly slow decay was an artifact of the time difference between the excitation turn-off timing and gate start time. The decaying part (delay time: 200--1000 ns) of the luminescence intensity was fit with a single exponential function considering the gate width. From the fitting, the paraexciton lifetime was determined to be 410 $\pm$ 8 ns. Further, we found no nonlinear decay, showing that two-body inelastic collision processes were absent from this density region.
The measured lifetime is compatible with the results of experiments at temperatures above 1 K.\cite{yoshioka2010quantum} Thus, there are no physical processes that shorten the lifetime at ultra-low temperatures below 1 K. In addition, the obtained lifetime supports our previous assumption of a 300-ns paraexciton lifetime to reproduce the paraexciton temperature of 100 mK.\cite{yoshioka2013generation} As discussed below, the lifetime is longer than the typical timescale associated with the trap potential and, thus, the present system is ideal for realizing the BEC.

\section{Exciton dynamics measurement}
Next, we measured the spatio-temporal dynamics of paraexcitons . We changed the excitation light intensity, its position relative to the trap potential, the ICCD camera gate setting, and the spectrometer slit width. 

To obtain space- and time-resolved luminescence spectra with a high signal-to-noise ratio, relatively strong emission from paraexcitons is required. Therefore, we set the excitation-light peak intensity to 30 mW to generate relatively high paraexciton density of $\sim$10$^{14}$ cm$^{-3}$. 

Secondly, to observe the paraexciton drift clearly, the drift distance must be larger than the detection spatial resolution ($\sim$10 $\mu$m). Therefore, the excitation position was intentionally shifted 170 $\mu$m from the bottom of the trap potential along the sample [100] axis. Paraexcitons generated at the position of the excitation drift to the bottom of the trap potential as shown in Fig. 4(a).

For the mobility and diffusion constant measurements, we focused on the buildup of the spatial distribution of paraexcitons. We varied the delay time between the ICCD gate pulse and the pulse for driving the AOM. We controlled the two pulses as shown in Fig. 4(b). To obtain sufficient signal intensity with a good signal-to-noise ratio, a long gate width is required. However, the gate width should be shorter than the trap frequency ($\sim$10 MHz) of the potential. Therefore, we set the gate width to 30 ns. We gradually varied the delay between the two timing pulses in 30-ns steps. The pulse repetition frequency was 500 Hz. The exposure time was typically set to 1 h. The mixing-chamber temperature under the optical excitation was 58 mK.

We observed not only the spatio-temporal dynamics, but also the time evolution of the energy relaxation of the paraexcitons in the trap potential. To achieve this, we set a narrow spectrometer slit width ($\sim$30 $\mu$m) to obtain the maximum energy resolution. The longitudinal direction of the slit was parallel to the drift direction. 

\begin{figure}
   \begin{tabular}{l}

      % 1
      \begin{minipage}{1\hsize}
       
      (a)  \includegraphics[width=4.0cm]{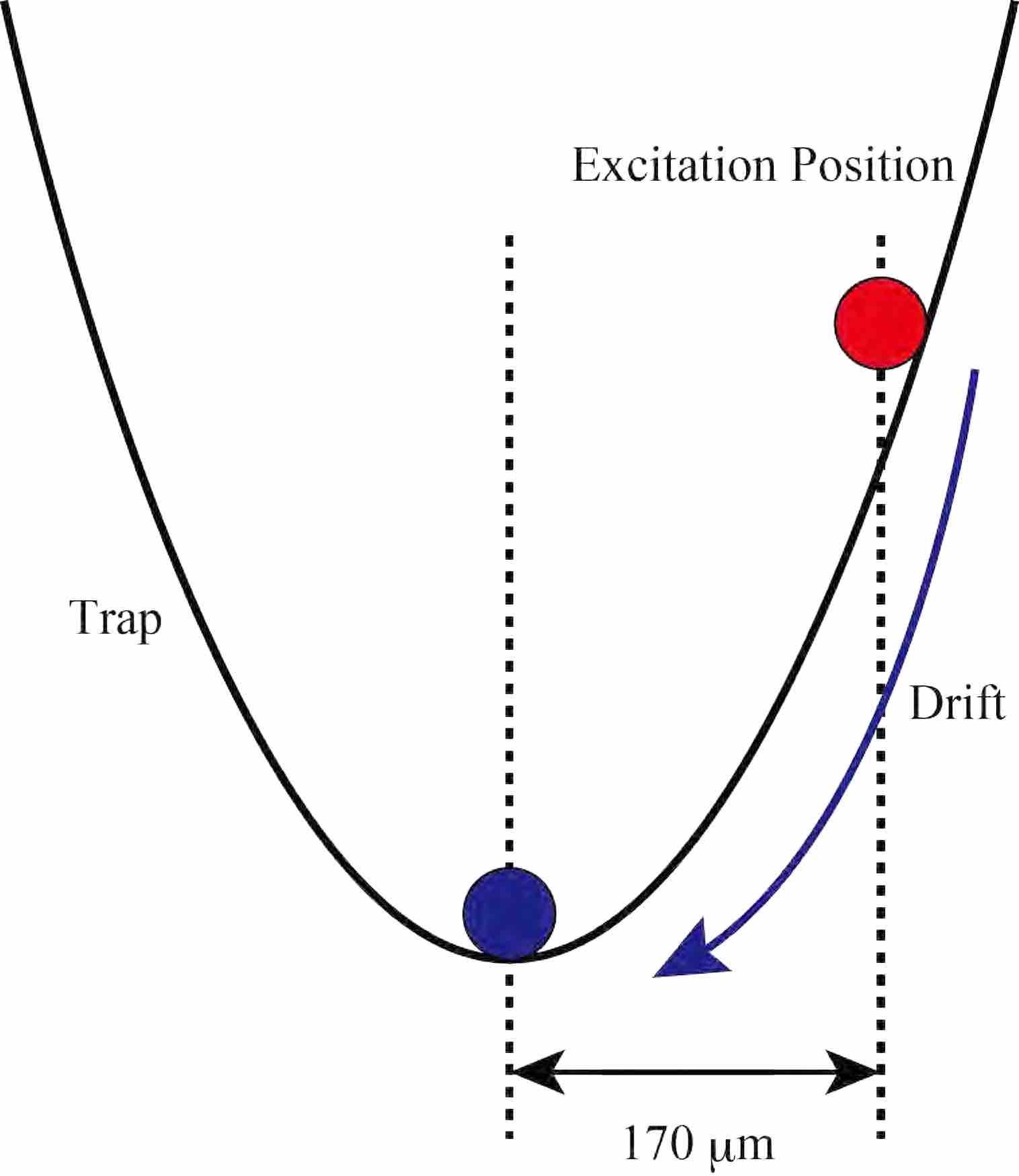}\hspace{0.1cm}
      
      \end{minipage}\\ \\

      % 2
      \begin{minipage}{1\hsize}
        
      (b)  \includegraphics[width=7.0cm]{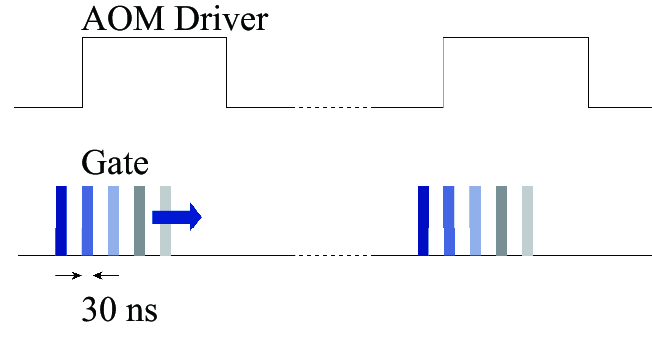}\hspace{0.1cm}

      \end{minipage}

    \end{tabular}

  \ \ \ \ \ \
 \caption{\label{fig3}
Settings for spatio-temporal dynamics measurements.
(a) Schematic illustration of optical excitation spatial configurations. The red circle shows the excitation position in the trap potential along the sample [100] axis. Paraexcitons drifted from the excitation position to the bottom of the trap potential (blue circle). The distance between the excitation position and the trap potential bottom was set to 170 $\mu$m. 
(b) Timing chart of two electric pulses for spatio-temporal dynamics measurements. (Upper row) Timing chart of AOM driver control pulse. The pulse duration varied with the excitation power. (Lower row) Gate pulses sent to ICCD. We varied the delay time between the control pulse and gate to observe the spatio-temporal dynamics. }
\end{figure}

\begin{figure}
    \begin{tabular}{c}

      % 1
      \begin{minipage}{0.5\hsize}
        \begin{center}
          \includegraphics[clip, width=4.3cm]{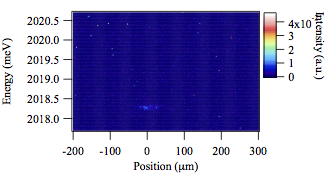}
          \hspace{1.6cm} (a)
        \end{center}
      \end{minipage}

      % 2
      \begin{minipage}{0.5\hsize}
        \begin{center}
          \includegraphics[clip, width=4.3cm]{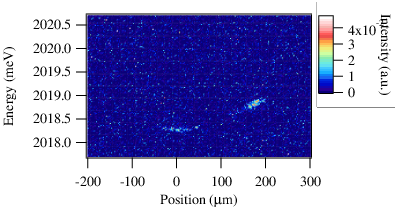}
          \hspace{1.6cm} (b)
        \end{center}
      \end{minipage} \\

      % 3
      \begin{minipage}{0.5\hsize}
        \begin{center}
          \includegraphics[clip, width=4.3cm]{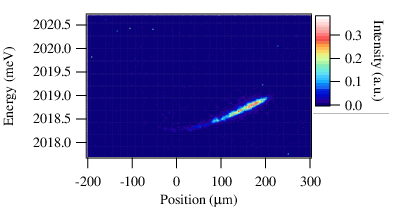}
          \hspace{1.6cm}(c)
        \end{center}
      \end{minipage}
      
      \begin{minipage}{0.5\hsize}
        \begin{center}
          \includegraphics[clip, width=4.3cm]{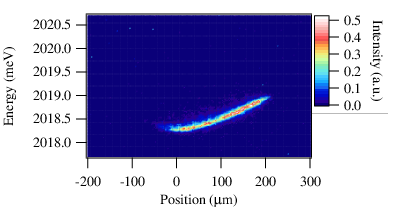}
          \hspace{1.6cm}(d)
        \end{center}
      \end{minipage}\\
      
      \begin{minipage}{0.5\hsize}
        \begin{center}
          \includegraphics[clip, width=4.3cm]{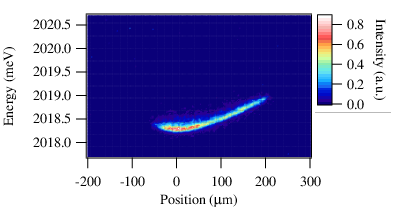}
          \hspace{1.6cm}(e)
        \end{center}
      \end{minipage}
      
      \begin{minipage}{0.5\hsize}
        \begin{center}
          \includegraphics[clip, width=4.3cm]{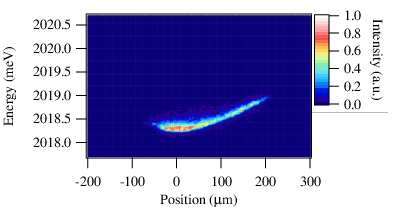}
          \hspace{1.6cm}(f)
        \end{center}
      \end{minipage}\\

	\begin{minipage}{0.5\hsize}
        \begin{center}
         \includegraphics[clip, width=6cm]{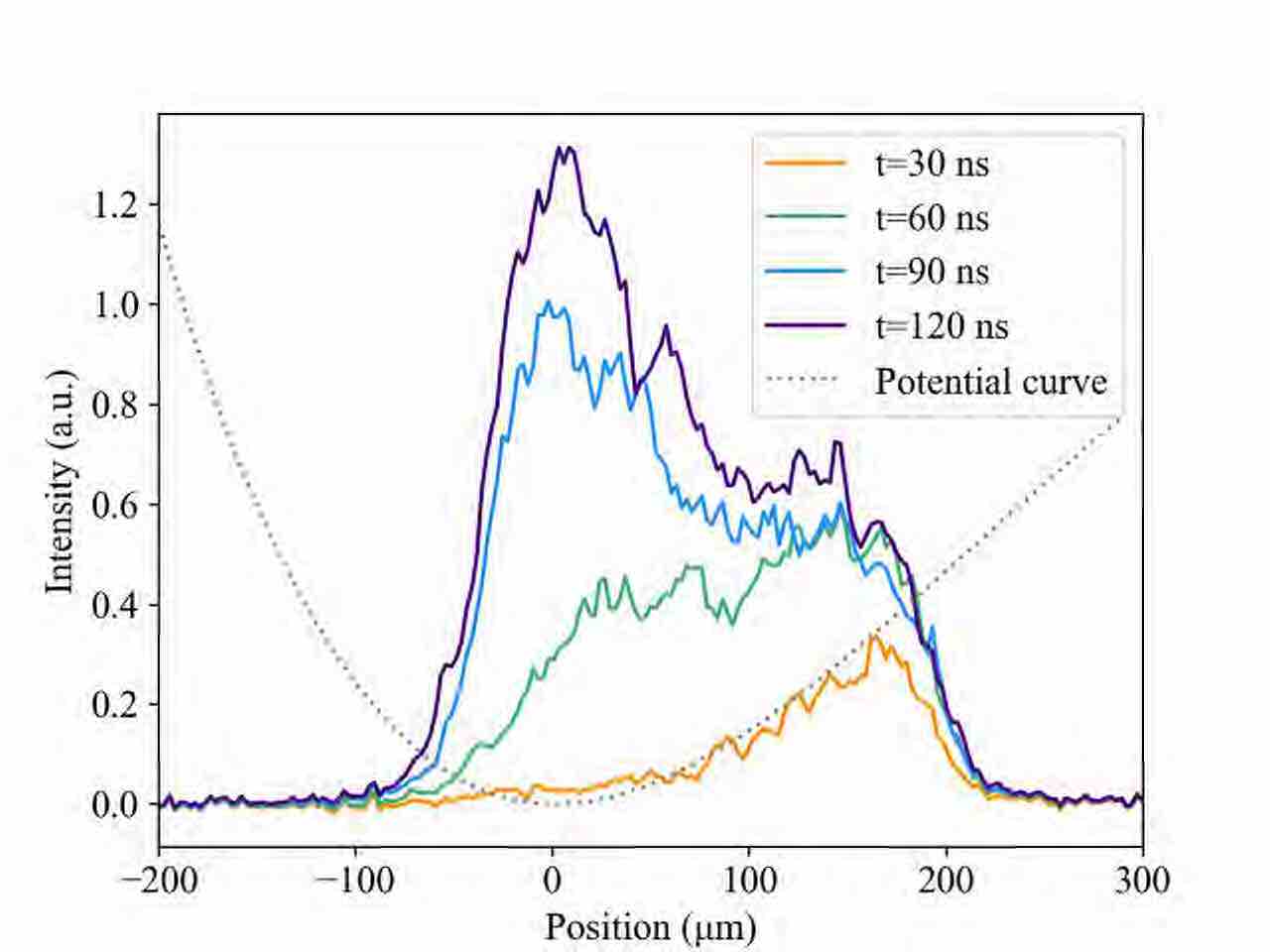}
         \hspace{1.6cm}(g)
        \end{center}
      \end{minipage}

    \end{tabular}
    \caption{ (a--f) Time evolution of spatially resolved luminescence spectra of paraexcitons after activation of excitation light for time $t$ of (a) --30, (b) 0, (c) 30, (d) 60, (e) 90, and (f) 120 ns. All signals in (a)--(e) were normalized with the peak intensity in (f).
(g) Spatially resolved luminescence intensity of paraexcitons at each $t$. The gray dotted line shows the trap potential shape. 
}
    \label{fig5}
  
\end{figure}

\begin{figure}
    \begin{tabular}{c}

      % 1
      \begin{minipage}{0.5\hsize}
        \begin{center}
          \includegraphics[clip, width=4.3cm]{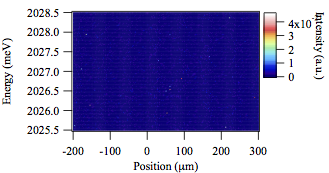}
          \hspace{1.6cm} (a)
        \end{center}
      \end{minipage}

      % 2
      \begin{minipage}{0.5\hsize}
        \begin{center}
          \includegraphics[clip, width=4.3cm]{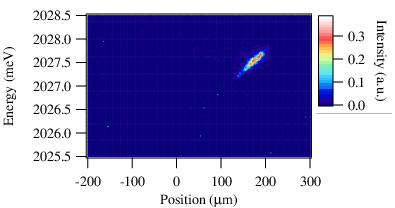}
          \hspace{1.6cm} (b)
        \end{center}
      \end{minipage} \\

      % 3
      \begin{minipage}{0.5\hsize}
        \begin{center}
          \includegraphics[clip, width=4.3cm]{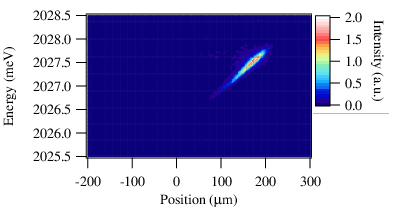}
          \hspace{1.6cm}(c)
        \end{center}
      \end{minipage}
      
      \begin{minipage}{0.5\hsize}
        \begin{center}
          \includegraphics[clip, width=4.3cm]{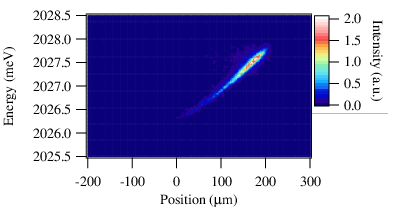}
          \hspace{1.6cm}(d)
        \end{center}
      \end{minipage}\\
      
      \begin{minipage}{0.5\hsize}
        \begin{center}
          \includegraphics[clip, width=4.3cm]{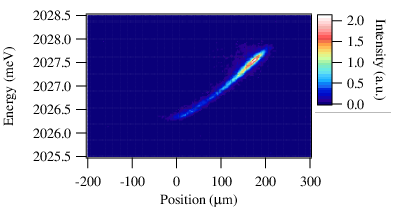}
          \hspace{1.6cm}(e)
        \end{center}
      \end{minipage}
      
      \begin{minipage}{0.5\hsize}
        \begin{center}
          \includegraphics[clip, width=4.3cm]{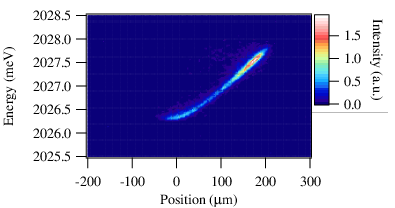}
          \hspace{1.6cm}(f)
        \end{center}
      \end{minipage}\\

    \end{tabular}
    \caption{ (a--f) Time evolution of spatially resolved luminescence spectra of orthoexcitons after excitation light activation, for $t$ of (a) --30 ns, (b) 0, (c) 30, (d) 60, (e) 90, and (f) 120 ns. All signals in (a)--(e) were normalized with the peak intensity in Fig. 5(f) (paraexcitons).}
    \label{fig5}
  
\end{figure}

We obtained six images each for the direct luminescence from the paraexcitons and orthoexcitons, as shown in Figs. 5(a--f) and 6(a--f). The time origin was set to the appearance point of the paraexciton luminescence.
The following three points were noted from the results: \\
(1)	When the orthoexciton luminescence signal appeared  ($t$ = 0 ns, Fig. 6(b)), the paraexciton luminescence was very weak (Fig. 5(b)). At the next time gate, the paraexciton luminescence appeared clearly. These results suggest that the time scale of the orthoexciton-to-paraexciton conversion process is of the same order as the gate width ($\sim$30 ns) at sub-Kelvin temperatures. This time scale seems different from the orthoexciton-to-paraexciton conversion time (a few nanoseconds) reported previously,\cite{weiner1983ortho,snoke1990mechanism,jang2004spin} at a 2-K lattice temperature. Further investigation is required to understand the physical origin of the discrepancy. \\
(2)	For $t=30$--90 ns, Figs. 5(c) and 5(e) show that the peak position of the paraexciton luminescence drifts to the bottom of the trap potential over 170 $\mu$m. Thus, we successfully observed paraexciton drift over a macroscopic range. \\
(3)	For $t=90$--120 ns, Fig. 5(e) shows that the peak position is at the trap potential bottom, and the overall spatial distribution is almost identical to that in Fig. 5(f). This result shows that those spatial distributions are of the paraexciton steady states, which are achieved within 120 ns, which is considerably shorter than the paraexciton lifetime. 

Furthermore, we estimated the trap frequencies from Fig. 5(e), as 13, 9, and 9 MHz in the directions of the [100], [011], and [01$\bar{1}$] axes, respectively. Hence, the applied force was expected to be 170 N. The trap frequencies were used for the numerical analysis reported in the following section.

In addition, an interesting orthoexciton characteristic can be deduced from Figs. 6(a--f). The overall spatial distribution of the orthoexciton luminescence changed between $t$ = 90 and 120 ns (Figs. 6(e) and 6(f)), although the overall spatial distributions of paraexcitons in Figs. 5(e) and 5(f) were almost identical. Our results suggest that the spatial distribution of orthoexcitons does not reach a steady state within the time scale of the orthoexciton-to-paraexciton conversion process, which dominates the orthoexciton lifetime. Furthermore, for t=120 ns, the orthoexciton density increases at the peak position of the paraexciton distribution. Therefore, this phenomenon implies that another source of orthoexciton generation appear after t = 90 ns, i.e., an up-conversion process from paraexcitons to orthoexcitons. This up-conversion process is expected to be an important topic for future studies of high-density paraexcitons.

\section{Numerical simulation}
To estimate the mobility and the diffusion constant of paraexcitons, we compared the results of the spatio-temporal dynamics measurements with numerical simulations of the paraexciton luminescence spatial distribution along the sample [100] axis. The mobility and diffusion constant values were used as fitting parameters in the numerical calculations. We examined the experimentally measured spatio-temporal dynamics for three discrete time delays with finite gate time, reported in the previous section. However, it was impossible to resolve the dynamics within the gate time. Therefore, for the calculation, we assumed that the diffusion constant and mobility were unchanged within the gate time and underwent step-like changes as the delay time proceeded, with $t$ values of 1) 0--30, 2) 30--60, and 3) 60--90 ns. We considered the diffusion and drift terms and the paraexciton decay and generation in the numerical calculations.
	The basic equation was based on the conventional diffusion equation used for numerical calculations. We observed the change of the luminescence intensity distribution along the sample [100] axis. To analyze this temporal evolution, we set a diffusion equation, which describes the spatio-temporal dynamics of the paraexcitons in one direction. The equation was expressed as
\begin{eqnarray}
\frac{\partial}{\partial t} N(z,t)&=&\frac{\partial}{\partial z}(\mu(T(z,t))\frac{\partial V(z)}{\partial z})N(z,t) \nonumber \\
&+&\frac{\partial}{\partial z} D(T(z,t)) \frac{\partial}{\partial z} N(z,t)\nonumber \\
&-&\frac{1}{\tau_p(z)}N(z,t)+G(z,t),
\end{eqnarray}
where $N(z,t)$ is the paraexciton density at position $z$ and time $t$, $\mu(z,T)$ denotes the paraexciton mobility at $z$ and the paraexciton temperature $T(z,t)$, $V(z)$ is the potential energy of the trap potential at $z$, 
$t_p(z,t)$ denotes the effective lifetime of paraexcitons at $z$ and $t$, and $G(z,t)$ is the paraexciton generation rate at $z$.
The static $V(z)$ was calculated independently according to the conventional Hertzian contact problem. $G(z,t)$ was calculated from the excitation light intensity, actual absorption, and excitation volume.
We calculated the effective lifetime while considering the two-body inelastic collision loss:
\begin{equation}
\frac{1}{\tau_p(z,t)}=\frac{1}{\tau_0}+A\cdot N(z,t),
\end{equation}
where $\tau_0$ and $A$ are the lifetime extracted in section 4 ($\tau_0$ = 410 ns) and the two-body collision loss coefficient (set to 10$^{-16}$ cm$^3$/ns),\cite{yoshioka2010quantum} respectively. 

We performed numerical calculations based on the above equation, as follows. \\
(1) Our diffusion equation was nonlinear because the effective lifetime is dependent on the density distribution. However, the density distribution time evolution within the gate time is unknown. Here, we evaluated the time-averaged density distribution as follows, and calculated the position-dependent effective lifetime:\\
For 0 $\leq t <$ 30 ns, 30 $\leq t <$ 60 ns, and 60 $\leq t <$ 90 ns, $N(z,t)= N_1(z,t)$ was estimated from the luminescence distributions obtained from our measurements at $t$ = 30, 60, and 90 ns, respectively. \\
(2)	We set the time and space grid as
\begin{eqnarray}
z_i,\quad \Delta z=1 \ \mathrm{\mu m},\quad i=1,2,..,500, \\
t_j,\quad \Delta t=0.5 \ \mathrm{ns},\quad j=1,2,..,180.
\end{eqnarray}
As a result of our approximation of the effective lifetime by the time-averaged density, the time evolutions of the distribution for paraexcitons generated at different times were independent. Therefore, summation of the density distribution at the end of the calculations was sufficient to obtain the final result. The density distribution calculation procedure was as follows:
We defined $n_m (z,t)$ as the density distribution of paraexcitons generated when $t=m\cdot \Delta t$. For $t< m \cdot \Delta t$, $n_m(z,t)=0$; for $t=m\cdot \Delta t$, $n_m(z,t)= G(z,t)\cdot \Delta t$; and for $t >m\cdot \Delta t$,
\begin{eqnarray}
n_m(z,t+\Delta t)=\frac{\partial}{\partial z}(\mu_m(z,t)\frac{\partial V(z)}{\partial z})n_m(z,t)\cdot \Delta t \nonumber \\
+\frac{\partial}{\partial z}D_m(z,t)\frac{\partial}{\partial z} n_m(z,t)\cdot \Delta t-\frac{\Delta t}{\tau_p}n_m(z,t)+n_m(z,t).
\end{eqnarray}
The total paraexciton density was calculated from
\begin{eqnarray}
N(z,\tau)=\sum^{m <\tau/\Delta t}_{m=0}n_{m}(z,\tau).
\end{eqnarray}\\
(3) The fitting parameters were the mobility and diffusion constant, which are not linearly independent. We assumed that these parameters were dependent on the paraexciton temperature, as shown in Ref. 22.
\begin{eqnarray}
\mu_m(z,t)=\mu_m(T_m(z,t)), \\
D_m(z,t)=D_m(T_m(z,t)),
\end{eqnarray}
where $T_m(z,t)$ is the temperature of paraexcitons generated when $t=m\cdot \Delta t$ at $z$ and $t$. Moreover, the temperature dependence of these parameters (as equations below shows) could be estimated from the stress applied to the crystal (2.1 kbar). This estimation was based on the results provided by Wolfe's group paper,\cite{trauernicht1986drift} which considers interaction between the paraexcitons and TA phonons. The interaction strength and parameter temperature dependence are stress-dependent. We could regard the stress as approximately constant in our observation area range.
\begin{eqnarray}
\mu_m(z,t)=k_\mu \cdot T_m^{-1.73}, \\
D_m(z,t)=k_D\cdot T_m^{-0.73},
\end{eqnarray}
where $k_\mu$ and $k_D$ are constants.

We assumed that the cooling process was independent of the interactions between paraexcitons. Therefore, the temperature of the paraexcitons generated when $m\cdot \Delta t$ did not depend on $N(z,t)$, but rather on $(t- m\cdot \Delta t)$. As mentioned above, in our analysis, we estimated the diffusion constant and mobility from comparison of three images with our numerical calculations. Therefore, we again assumed average values of the paraexciton temperature, diffusion constant, and mobility corresponding to three time regions (t = 0--30 ns, t = 30--60 ns, t = 60--90 ns).:\\
For 0 ns $\leq t-m\cdot \Delta t<$ 30 ns,
\begin{eqnarray*}
T_m(t)=T_1,\ \mu_m(z,t)=k_\mu \cdot T_1^{-1.73},\\ D_m(z,t)=k_D\cdot T_1^{-0.73};
\end{eqnarray*}
For 30 ns $\leq t-m\cdot \Delta t<$ 60 ns,
\begin{eqnarray*}
T_m(t)=T_2,\ \mu_m(z,t)=k_\mu \cdot T_2^{-1.73},\\ D_m(z,t)=k_D\cdot T_2^{-0.73};
\end{eqnarray*}
For 60 ns$\leq t-m\cdot \Delta t<$ 90 ns,
\begin{eqnarray*}
T_m(t)=T_3,\ \mu_m(z,t)=k_\mu \cdot T_3^{-1.73},\\ D_m(z,t)=k_D\cdot T_3^{-0.73}.
\end{eqnarray*}\\
(4)	The signal detected by the ICCD camera was the integrated luminescence intensity during the gate time. To compare the numerical calculation and experiment results, integration of the varying density distributions was required. Moreover, the efficiency of the paraexciton luminescence was quadratically dependent on the stress \cite{kreingold1975investigation} at the paraexciton position and temperature. This temperature dependence is based on our model,\cite{yusuke_to} considering the local distribution function and momentum of the photon emitted through direct recombination of paraexcitons. Therefore, we calculated the integrated luminescence intensity as follows:
\begin{eqnarray}
I_1(z)\propto\sum^{t=30\ \mathrm{ns}}_{t=0\ \mathrm{ns}}(\sum_{m=0}^{m=180} n_m(z,t)\cdot \sigma(z)^2 \cdot \eta(T_m(t))), \\
I_2(z)\propto\sum^{t=60\ \mathrm{ns}}_{t=30\ \mathrm{ns}}(\sum_{m=0}^{m=180} n_m(z,t)\cdot \sigma(z)^2 \cdot \eta(T_m(t))), \\
I_3(z)\propto\sum^{t=90\ \mathrm{ns}}_{t=60\ \mathrm{ns}}(\sum_{m=0}^{m=180} n_m(z,t)\cdot \sigma(z)^2 \cdot \eta(T_m(t))),
\end{eqnarray}
where $\sigma(z)$ is the strain at $z$ and $\eta(T)$ is the luminescence efficiency at $T$:
\begin{eqnarray}
\eta(T)=T^{-1.5}\cdot \exp{(-\frac{(\hbar k_l)^2}{2m_{para}}\cdot\frac{1}{ k_BT})},
\end{eqnarray}
where $k_l$ is the wavenumber of the photon emitted through paraexciton direct recombination and $m_{para}$ is the paraexciton effective mass.

\section{Discussion}
We fit the results of our numerical calculations to each spatially resolved luminescence intensity, which revealed the paraexciton spatio-temporal dynamics. The fitting results are shown in Fig. 7. We adopted the diffusion constant and mobility as fitting parameters; these properties were estimated as functions of time, as shown in Figs. 8(a) and 8(b). These figures show that these values increased with time evolution. As expected, the increasing tendency implies that the impurity scattering or local potential fluctuations do not affect the transport properties of paraexcitons below 1 K. The diffusion constant was found to be 650, 850, and 1230 cm$^2$/s at $t$ = 30, 60, and 90 ns, respectively. The mobility was found to be 1.1, 2.1, and 5.1 $\times$ 10$^7$ cm$^2$/V$\cdot$s at $t$ = 30, 60, and 90 ns, respectively. The mobility value obtained here is the highest yet reported for excitons, to the best of our knowledge. 
\begin{figure}
 \includegraphics[width=8.6cm ]{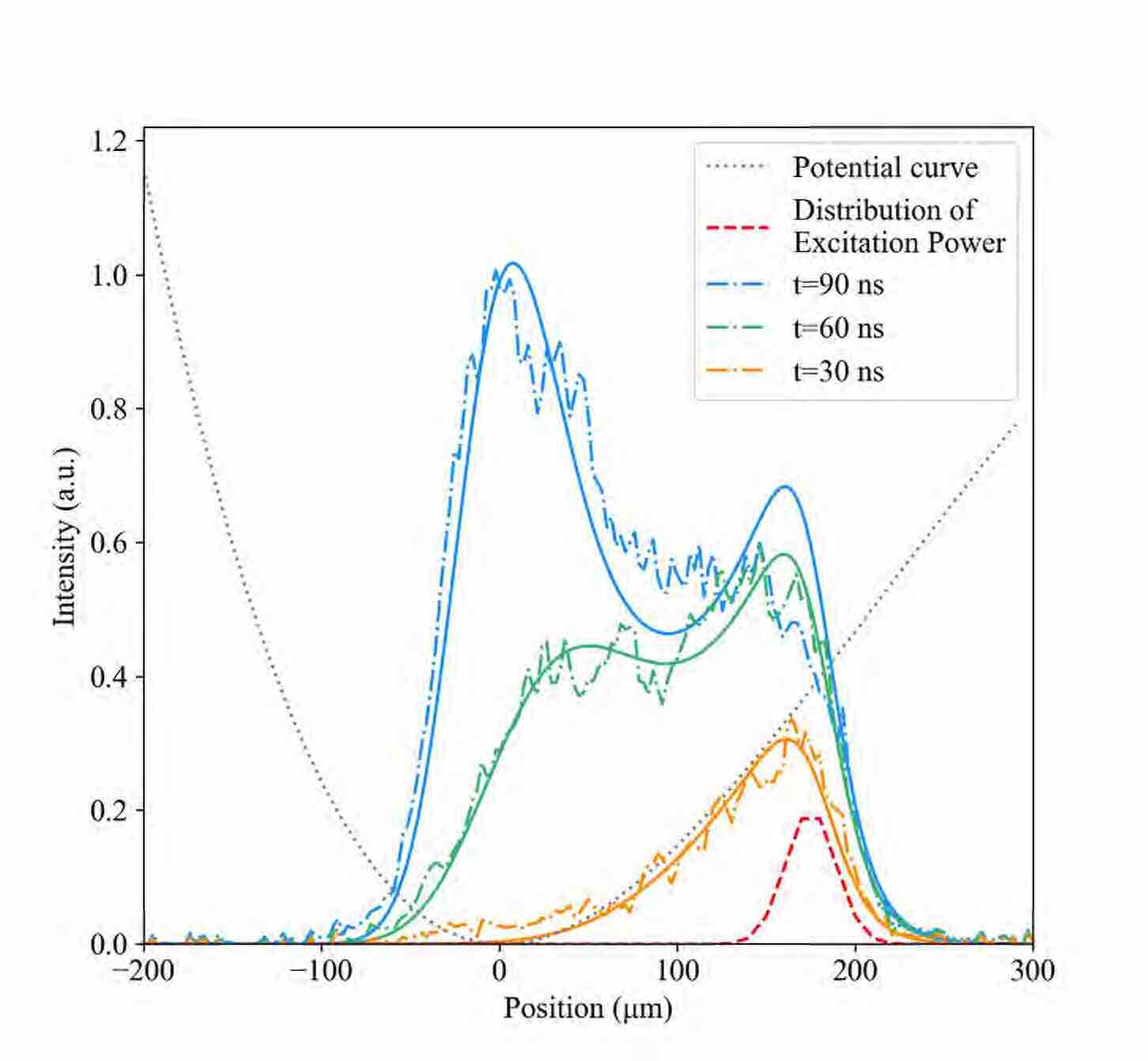}
\caption{\label{fig6}
Comparison of experiment and numerical simulation results. 
The gray and red dotted curves show the normalized trap potential shape and normalized excitation-light intensity distribution, respectively. The orange (green and blue) dashed curve shows the luminescence distribution at $t$ = 30 ns (60 and 90 ns) obtained in our experiment. The orange (green and blue) solid curve shows the simulated spatial distribution at $t$ = 30 ns (60 and 90 ns). 
}
\end{figure}

\begin{figure}
(a)\includegraphics[width=6.0cm ]{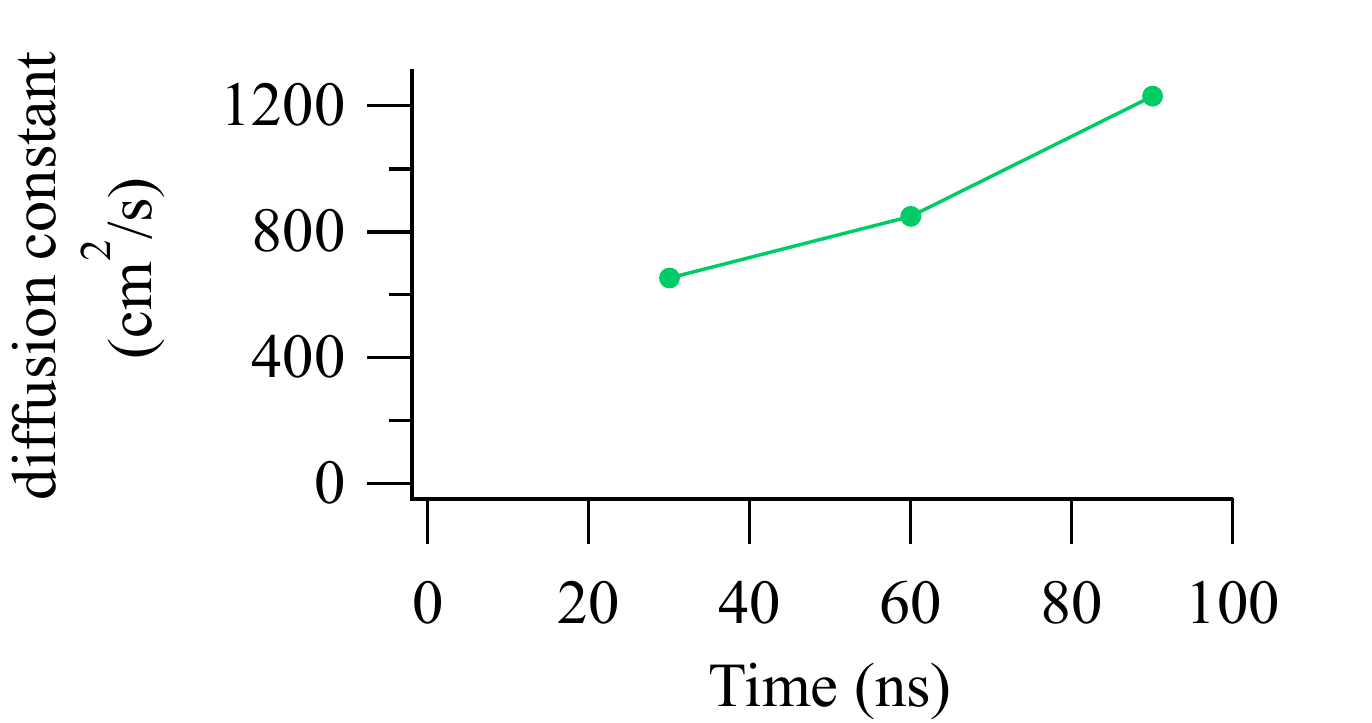}\\
(b) \includegraphics[width=6.0cm ]{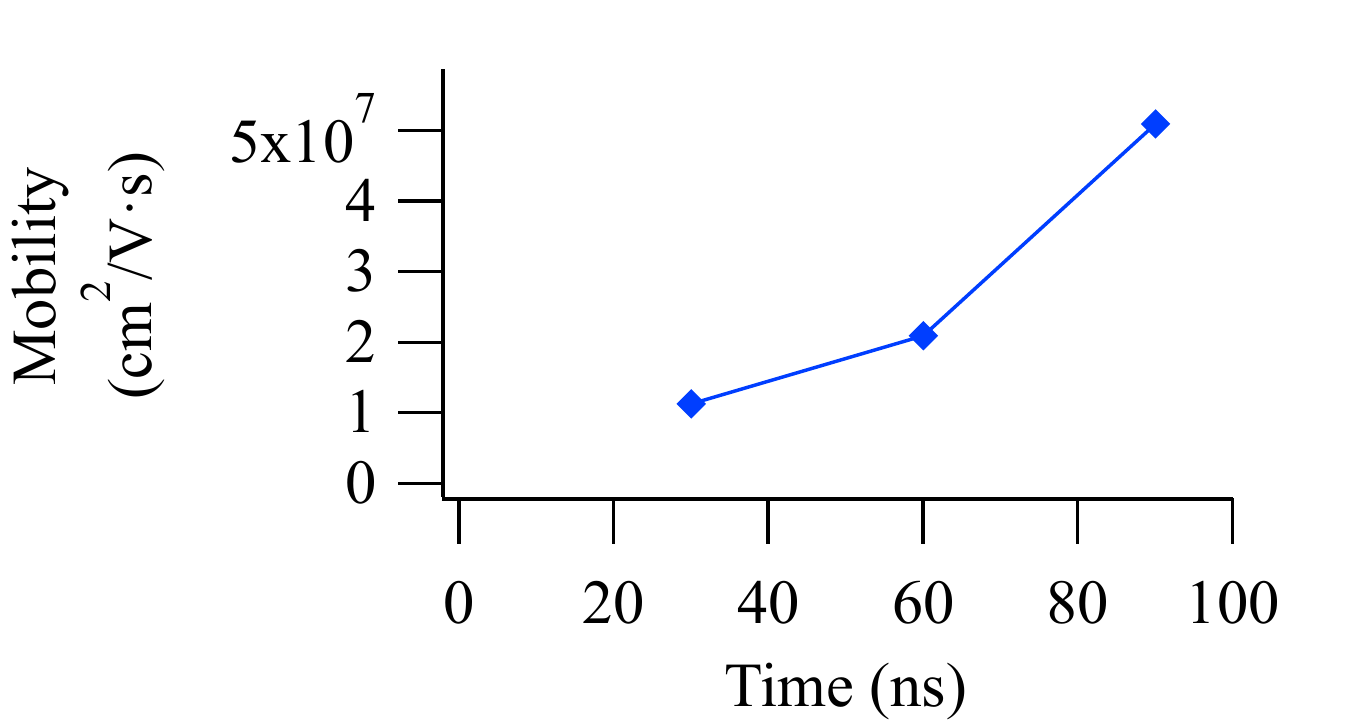}

\caption{\label{fig6b}
Diffusion constant and mobility as functions of time. (a, b) The green (blue) points show the diffusion constant (mobility) value at each $t$, extracted from our numerical calculations.  
}
\end{figure}

To examine the results, we calculated the mean free path $L$ of paraexcitons from the diffusion constant and corresponding the paraexciton temperature $T$. The paraexciton temperature at each delay time was extracted as a parameter for the diffusion constant and mobility from the fitting. 
\begin{eqnarray}
D=\frac{1}{3}v_{\mathrm{th}}L, \
v_{\mathrm{th}}=\sqrt{\frac{3k_BT}{m}}.
\end{eqnarray}
Hence, we could evaluate $L$ from the paraexciton diffusion constant and thermal velocity, yielding $L=$ 300 $\mu$m ($t=90$ ns.) Thus, paraexcitons can travel over the macroscopic range without collisions between excitons and phonons after $t=90$ ns. 
We also calculated the mean collision time, which increased over time. The estimated mean collision time at $t= 90$ ns was 400 ns, almost identical to the paraexciton lifetime of $\sim$410 ns. Thus, paraexcitons that reached the trap potential bottom after $t=90$ ns experienced only approximately one collision within their remaining lifetime.

\begin{figure}
 \includegraphics[width=7.0cm ]{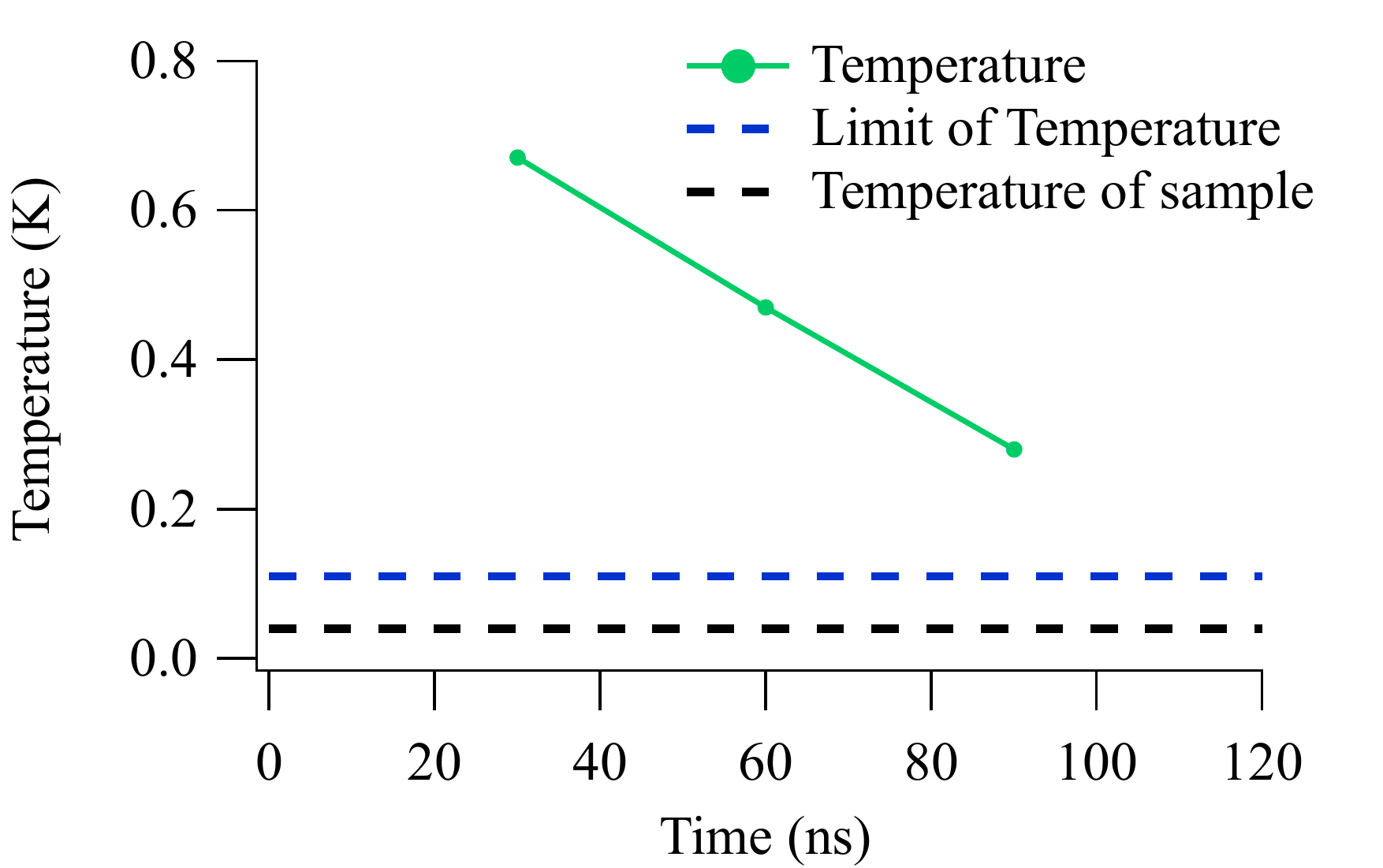}
\caption{\label{fig6d}
Time evolution of paraexciton temperature $T$. The green points and line show the paraexciton $T$ time evolution, estimated at each $t$ from numerical calculations. The blue and black dashed lines show the 110-mK temperature of steady-state paraexcitons at 2.1 kbar estimated from our prior paper\cite{yoshioka2013generation} and the mixing-chamber temperature of 58 mK, respectively.
}
\end{figure}

We next comment on the thermal relaxations of paraexcitons. As explained above, the paraexciton temperature at a given time was estimated by setting it as the parameter for diffusion constant and mobility from the fitting. The paraexciton temperature decreased during the drift to the trap potential bottom, as shown in Fig. 9. However, $T$ does not reach below $\sim$100 mK as reported in our prior study.\cite{yoshioka2013generation} This implies that, after $t= 90$ ns, the paraexciton cooling efficiency is limited and confirms that the predominant cooling mechanism at very low temperature is the paraexciton--TA-phonon scattering, as mentioned in our previous study.\cite{yoshioka2013generation} In the previous paragraph, we reported that, after $t= 90$ ns, paraexcitons experience approximately one collision within their lifetime. This result is consistent with the observed cooling process limitation after $t= 90$ ns.

Next, we checked how the experimentally measured spatio-temporal dynamics differ from expectation based on the conventional temperature region in the vicinity of 1.2 K. We simulated the spatio-temporal dynamics with the mobility and diffusion constant at 1.2 K, as shown in Fig. 10. The results show that the majority of the created paraexcitons could not reach the trap bottom at 1.2 K, although they do achieve this at dilution-refrigerator temperatures. This is because the paraexciton--LA-phonon scattering process was still active and limited the paraexciton mobility in the trap potential. This finding indicates that the spatial steady state at 1.2 K clearly differs from the expected distribution assuming thermal equilibrium in the trap potential. Therefore, it is important to cool paraexcitons well below 1 K to study the quantum statistical effects of accumulated paraexcitons at the trap potential bottom.

\begin{figure}
 \includegraphics[width=8.6cm]{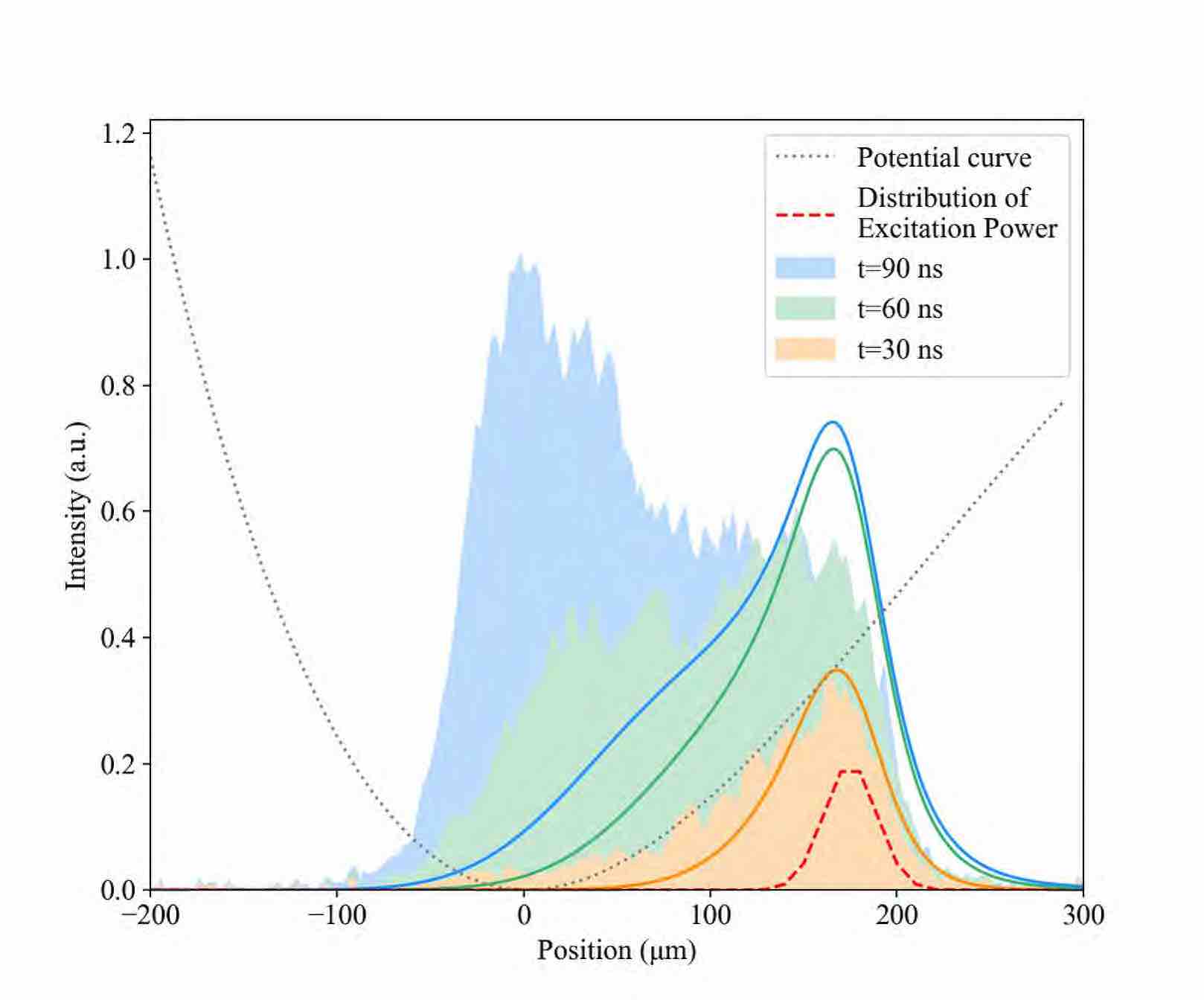}
\caption{\label{fig7a}
Simulated spatial distribution with fixed diffusion constant and mobility at 1.2 K.
The gray and red dotted lines show the trap potential shape and normalized excitation-light intensity distribution, respectively. The orange (green and blue) solid curve shows the simulated luminescence distribution at $t$ = 30 ns (60 and 90 ns). For comparison, the experimentally measured spatially resolved luminescence intensities at 58-mK lattice temperature are indicated by the filled regions.
}
\end{figure}
It is important to determine whether the spatial distribution of the 1s paraexcitons reached the distribution defined by the statistical distribution function and the three-dimensional trap potential shape only. Thus, we simulated the time evolution of the paraexciton distribution, as shown in Figs. 11(a) and 11(b), under the following conditions.\\
(1)	We tracked the distribution of paraexcitons generated by a pulsed excitation light with 1-ns duration to avoid overlap of the distributions of paraexcitons generated at various times. \\
(2)	The time evolution of the paraexciton temperature and the transport properties confirmed to the fitting results. That is, the diffusion constant was 650, 850, and 1230 cm$^2$/s for $t=0$--30, 30--60, and 60--90 ns, respectively; and the mobility was 1.1, 2.1, and 5.1 $\times$ 10$^7$ cm$^2$/V$\cdot$s for $t=0$--30, 30--60, and 60--90 ns, respectively. Moreover, $T$ was 670, 470, and 280 mK at $t=0$--30, 30--60, and 60--90 ns, respectively. 
We found that the distribution peak arrived at the trap potential bottom well within the paraexciton lifetime. Furthermore, the distribution spatial width was in accordance with that defined by the statistical distribution function and the trap potential shape within 90 ns. Note that a period of only $\sim$15 ns was required for the spatial width to accord with that defined by the statistical distribution function and the trap potential shape when $T$ changed abruptly from 470 to 280 mK. Therefore, we can expect that, after $t=90$ ns, the spatial width almost instantaneously accords with that defined by the statistical distribution function and the trap potential shape, as the paraexciton $T$ decreases. This is a unique feature associated with paraexcitons in the 100-mK temperature region. 

\begin{figure}
 \includegraphics[width=4.0cm]{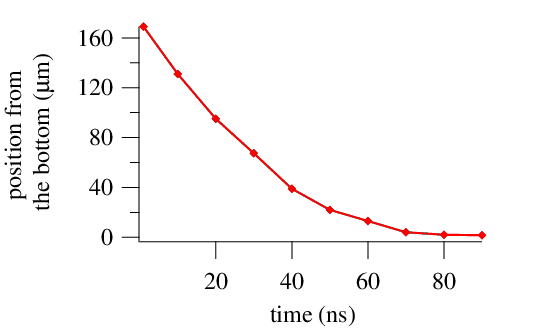}
  \includegraphics[width=4.0cm]{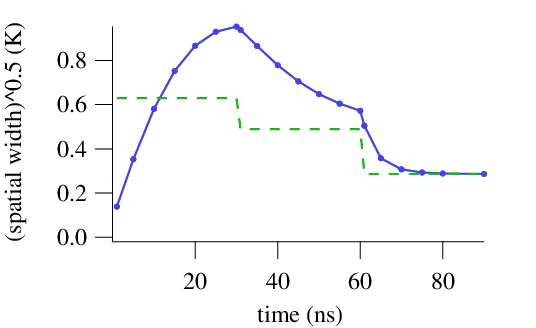}
\caption{\label{fig7b}
Simulated paraexciton distributions generated by short pulse. The excitation pulse width was 1 ns and the time origin was set to the paraexciton generation time.
(a) The red circles and solid line show the peak position drift of the distribution. (b) The blue circles and solid line show the time evolution of the spatial distribution width. The vertical axis shows the paraexciton $T$ estimated from the spatial width (Full width at half maximum). The green dashed line shows the assumed time evolution of $T$, which we adopted as the simulation parameter.
}
\end{figure}

As explained in the previous section, we considered the two-body inelastic collision loss when performing the numerical calculations, with the corresponding loss coefficient for the fitting of $\sim10^{-16}$ cm$^{-3}$/ns. This value was reported previously.\cite{yoshioka2010quantum} We examined the influence of the two-body inelastic collision coefficient on the fitting results, by fitting the experimental results with different two-body inelastic collision loss coefficient values (10$^{-15}$ and 10$^{-17}$ cm$^{-3}$/ns), as shown in Figs. 12(a) and 12(b). These results show that the goodness of fit is clearly dependent on the two-body inelastic collision loss coefficient. The goodness of fit in Fig. 7 is better than that in Figs. 12(a) and 12(b) around the bottom of the trap. Therefore, the adopted coefficient value is reasonable. The coincidence of the collision coefficient with those reported in previous works \cite{yoshioka2010quantum} suggests that the two-body inelastic collision coefficient is also temperature-independent in the 100-mK temperature region.

\begin{figure}
 (a)\includegraphics[width=8.0cm]{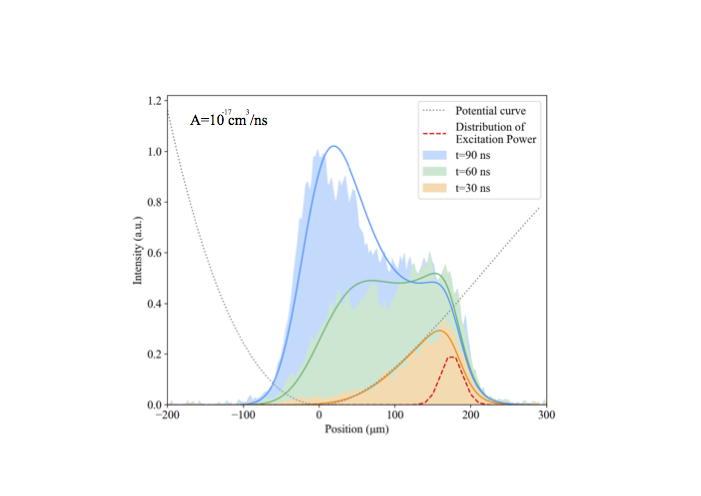}\\
(b)\includegraphics[width=8.0cm]{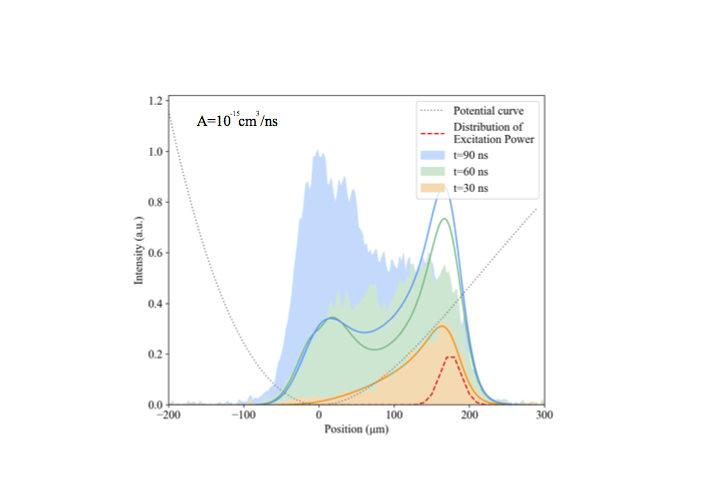}
\caption{\label{fig7d}
Simulated spatio-temporal dynamics with two-body inelastic collision rates of (a) 10$^{-17}$ and (b) 10$^{-15}$ cm$^{-3}$/ns.
The gray and red dotted lines show the trap potential shape and the normalized excitation light intensity distribution. The orange (green and blue) solid curve shows the calculated luminescence distribution at $t$ = 30 ns (60 and 90 ns). The experimentally measured spatially resolved luminescence intensities are shown by the filled regions.
}
\end{figure}

\section{Conclusion}
In conclusion, we observed basic parameters such as the lifetime, mobility, and diffusion constant of paraexcitons in a strain-induced trap at very low temperatures below 1 K, which were achieved using a dilution refrigerator. These parameters are important because they determine the paraexciton spatio-temporal dynamics, and were obtained by observing the space- and time-resolved luminescence spectrum of the paraexcitons in the trap potential. We extracted a 410-ns lifetime from measurements of the luminescence intensity decay. Moreover, we observed the buildup of the paraexciton spatial distribution and compared this spatio-temporal dynamics with the results of numerical calculations. We extracted the diffusion constant and mobility ($\sim$10$^2$ cm$^2$/ns and $\sim10^7$ cm$^2$/V$\cdot$s, respectively), which were found to increase with time. In particular, we obtained mobility of 5.1 $\times\,10^7$ cm$^2$/V$\cdot$s at a corresponding temperature of 280 mK. This value is the highest exciton mobility reported to date, to the best of our knowledge.

From these analyses, we found that there is no physical process that shortens the lifetime at ultra-low temperatures below 1 K. We also revealed that impurity scattering and local potential fluctuations have no effect on the paraexciton transport over 170 $\mu$m below 1 K, from the temperature dependence of diffusion constant and mobility.

One of the aims was to determine whether the spatial distribution of 1s paraexcitons reaches the distribution defined by the statistical distribution function and the three-dimensional trap potential shape. From our analyses, we found that ultra-low temperatures of paraexcitons well below 1 K, along with a paraexciton density that is not excessively high, are required to obtain the spatial distribution of paraexcitons defined by the statistical distribution function and the shape of the three-dimensional trap potential. We found that 1s paraexcitons can be treated as ideal Bose particles in the trap potential at low temperatures below 300 mK. Condensation of distributions in both momentum space and real space is expected at the paraexciton BEC transition, as for atomic BEC. Therefore, further experiments on high-density excitation of paraexcitons at low temperatures of approximately 100 mK are valuable as regards realization of the exciton BEC. Our experiments for realization of the exciton are currently on-going.

We performed time-resolved luminescence measurements at low temperatures below 1 K. Therefore, further experiments can be conducted. For example, the conversion process from orthoexcitons to paraexcitons in Cu$_2$O, as already briefly discussed in this paper, is an important subject of study. The conversion rate at temperatures above 2 K has been thoroughly studied, and several theoretical suggestions have been made as regards the conversion process mechanism.\cite{weiner1983ortho,snoke1990mechanism,jang2004spin} However, the conversion rates of the different theoretical models have different temperature dependencies, and experiments are limited at temperatures above 2 K. To identify the model that explains the conversion process mechanism, further study of the conversion rate at temperatures below 1 K will be helpful. Thus, our time-resolved luminescence spectroscopy using a dilution refrigerator may offer a unique opportunity for direct measurement of the conversion rate at the low temperatures.

Furthermore, measurement of spatio-temporal dynamics below 1 K may contribute to an existing controversy. Specifically, experimental observations of the exciton superfluidity could be addressed. Fortin et al. \cite{fortin1993exciton} studied the spatio-temporal dynamics of excitons in a strain-free Cu$_2$O crystal using photovoltaic measurements. They found that the time-resolved photovoltaic signal at 1.85 K changes from diffusive to ballistic motion with increased  density of paraexcitons, and attributed this sudden change to realization of the exciton superfluidity. However, other researchers have claimed that the phonon wind model can explain this change. \cite{kopelevich1996exciton} We note that the two models of the change from diffusive to ballistic motion differ significantly with regard to the temperature dependence of the critical excitation intensity at which the change occurs. Therefore, study of the spatio-temporal dynamics at ultra-low temperatures below 1 K may extend the temperature region to allow testing of these models, thereby revealing which best describes the peculiar spatio-temporal dynamics.
\section*{ACKNOWLEDGEMENT}
We thank E. Chae for valuable comments. This study was supported by a Grant-in-Aid for Scientific Research on Innovative Area “Optical science of dynamically correlated electrons (DYCE)” 20104002, by a Grant-in-Aid for Scientific Research from the Ministry of Education, Culture, Sports, Science, and Technology; JSPS KAKENHI (grant numbers 26247049, 25707024, JP17H06205 and 15H06131); by the Photon Frontier Network Program of the Ministry of Education, Culture, Sports, Science and Technology (MEXT), Japan, and by JSPS through its FIRST Program. 
\bibliographystyle{apsrev4-1}
\bibliography{arxiv_morita} %hoge.bibから拡張子を外した名前

\end{document}